# Enhanced Multigradient Dilution Preparation

Meenakshi Sanyal[1], Somenath Chakraborty[2]

[1] Saroj Mohan Institute of Technology, India
[2] The University of Southern Mississippi, USA
[1]meenakshi.sanyal2012@gmail.com,
[2]Somenath.Chakraborty@usm.edu

**Abstract**: In our paper the new algorithm enhanced multi gradient Dilution Preparation (EMDP) is discussed. This new algorithm is reported with a lab on chip or digital Microfluidic biochip to operate multiple operation on a tiny chip. We can use Digital Microfluidic biochip to operate multiple operation on a tiny chip. Samples are very costly which are used in any Biochemical laboratory Protocols. For the case of fast and high throughput application, It is essential to minimize the cost of operations and the time of operations and that is why one of the most challenging and important phase is sample preparation. In our proposed algorithm, we have hide to reduce sample droplets and waste droplets and for this purpose waste recycling is used, when different series of multi gradient targets concentration factors (CFS) are generated. We have compared our proposed algorithm with recent dilution techniques such as MTC, REMIA, and WARA. For the storage of intermediate droplets which, and generated during this process, on chip storage space $0(n)$ is needed.

## 1. Introduction

Digital micro fluidic (DMF) is biochip or 'Lab On Chip (LOC)' is one of the most emerging field of research in VLSI technology that is very large scale Integration Technology. In " More than Moore " technical trends, for the case of diabetes, cardiovascular diseases , DMF biochips helps to reduce the health care cost. Compare to all biochemical laboratory protocols, DMF biochips are faster.

Current flows in the Biochip for routing the micro fluidic droplets in biochip while in general information processing current does flow in the integrated circuits.

In our proposed algorithm, we have prepared multimode target dilution with minimum number of waste droplets and sample droplets. Sample droplets are very much costly and that is why, in our proposed algorithm, we have tried to minimise It for any number of dilution gradients.

Digital micro fluidic biochip is also essential in many real life application, drug discovery is one of them. With a certain concentration, we can charge the antibiotic, so that we can kill the bacteria. But if we use the antibiotics excessively, them some biological disorder can be occurred in human body. So for avoiding the effect of toxicity caused by various kinds of biochemical, it is essential to apply the popper amount of drug. So for generating the exact concentration of antibiotics, we have to apply the laboratory process for generating the multigradient dilution preparation. But this process takes a lot of time as well as to generate multi gradient dilution, more reagent is needed.

In our proposed algorithm, for preparing the multi gradient dilution, we have discussed the linear, harmonic, logarithmic and parabolic series. In the DMF biochip, our main objective to reduce the usage the sample droplets, because in most of the cases, they are very costly, and our another objective is to optimize the time and which is possible only with biochips. A generic algorithm, Mitra ET AL proposed, to prepare the multi gradient dilution for digital biochip, a generic algorithm. To reduce the usage of sample droplets, waste droplets and number of mix-split steps, all the efforts are given.

The rest of the paper is organized by the following process. In section 2, the basic idea of DMF biochip is discussed.

The architecture of the DMF Biochip cells, the mixing and splitting of the droplets are discussed in section 3, We have discussed the review work . The proposed algorithm is discussed in Section 4. In section 5, we have discussed the result and comparisons. At the end in section 6, the conclusion has been drawn.

## 2. Literature Review:

Recent years, Digital microfluidic supported biochips are widely used in many different kinds of fields like clinical diagnostics , biology manufactory, environmental monitor, analytical chemistry and military [1-3]. Digital microfluidic based biochips have received considerable attention as a promising platform for lab-on-a chip [4] realizations. Such composite systems manipulates fluids at a nanolitre or at picolitre volume scale[5]. By this procedure , the digital microfluidic based biochips simplify cumbersome laboratory procedures.
In various platforms, this systems can be used. For real- time bio-molecular detection ,and recognition [6],for massively parallel DNA Analysis, and for faster & cheaper clinical diagnosis , digital microfluidic based biochips are widely used. This systems can be used for immediate point-of-care health services, for the management of bio-terrorism threats, and for real time environmental toxicity monitoring [4]. In the chip design, the placement digital microfluidic based biochips is one of the key points. The physical potion of each biochemical analysis operation can be found with the smallest biochip area and the shortest completion, through the digital microfluidic based biochips[7-9].

## 3. Review Work:

In this century, various kinds of sample preparation have been established. WARA (Waste Recycling Algorithm),REMIA (Reactant Minimization Algorithm), and MTC (Multi Target Concentration )algorithm are most valuable among them .The first algorithm is REMIA .For the preparation of sample ,it reduces the usage of the reactant .The skewed mixing tree generation method is the main idea of REMIA . For "More and Moore "sample minimization, according to this method , the source droplet with much more higher concentration will be kept closer to the root . To explain the ex-potential dilution phase ,REMIA uses Exponential dilution trees. In this method , the concentration value of the droplet is denoted by each node of exponential dilution trees. In this method from the source droplet to the target droplet concentration one directed edge is pointed. And the concentration value of the root node is 1 . Since two resultant droplets are generated for exponential dilution, so for this case the out degree is 2 . In this method, one buffer droplet is needed for each operation.

The exponential dilution operation is represented by each branch note **.** Therefore the number of the dilution operation is implied by the total number of branch node. In WARA, It focused mainly in waste recycling for waste as well as sample minimization. The interpolated dilution process by divided by WARA into 3 phases **:-**

1. Mixing Tree generations
2. Droplet sharing
3. Droplet replacement

Now a days a proposed algorithm for multi target concentration. To ensure the highest optimization in number of mix- split steps, number of waste droplets and the number of sample droplets, a dilution graph based approach has been used.

## 4 Proposed work:

### 4.1 Algorithm:

In our proposed algorithm, EMDP ,here sample droplets (100%) dilutes buffer (0%)and recycles the waste droplets for generating the required CF. Here in our proposed method , we divides the dilution process into 3 steps.

1) Generation of serial dilution tree.
2) Creation of droplets with required CF.
3) Generating of the target series.

### 4.2 Generating serial dilution tree:

We construct Serial Dilution Tree(SDT) by the process serial dilution 1:1 mixing model is followed by the Serial Dilation Tree(SDT). The nodes of the SDT, represents the concentration factor of the droplets. The root of the SDT is the source droplets . Firstly the SDT involves one root droplet and one buffer droplet and it produces two identical droplets. Among them one droplet for next dilution and the another droplet for storage. We have mentioned the method in the following fig.

### 4.3 Droplet creation:

Here, in use proposed algorithm, we will create the required CF by diluting the immediate higher concentration droplet with buffer droplet or any other immediate droplet.

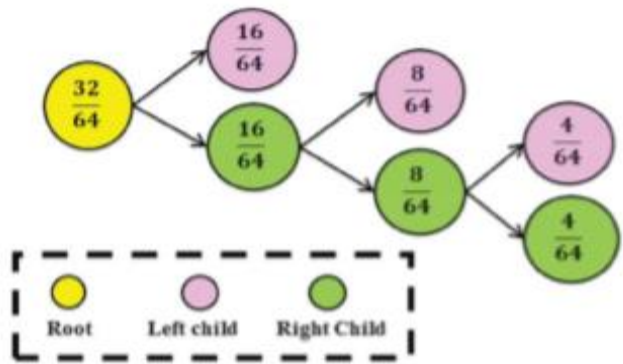

**Fig. 1.** Droplet Creation

### 4.4 EMDP(Enhanced Multigradient Dilution Preparation):

**Algorithm 1:**

*Serial Dilution Tree (root, target)*

**Input:** Source droplet and Target droplet concentration, sufficient buffer droplet

**Output:** A set of CFs

1. If root concentration = sample concentration
2. Number of sample = number of sample + 1
3. End if
4. If (root/2) >= target
5. Create two child note using mix and split(root,0)
6. Insert left child to storage
7. Number of buffer = number of buffer + 1
8. End while
9. Insert source into storage; circular queue

**Algorithm 2:**

*Create Droplet*

1. Let least higher CF value be stand into value and initially Value = 0.
2. While value = 0.
3. Search immediate higher concentration in the storage for required CF. Value=resultant droplet CF
4. If search is unsuccessful.
5. SDT (Sample concentration, required CF)

6. Else if value > (2 * CF of required droplet ) + 1/2^n.
7. Generate series (resultant droplet CF, required CF).
8. Value 0.
9. End if.

## Algorithm 3:

**CF Search**

**1**. do until all CF's are generated.

**2**. if CF value = sample concentration.

**3**. number of sample = number of sample + 1.

**4**. end if

**5**. Search the CF value.

**6**. if search is unsuccessful, continue with next CF

**7**. else

**8**. create droplet

**9**. end if

**4.5 Dilution Trees:**

Dilution Tree 1:

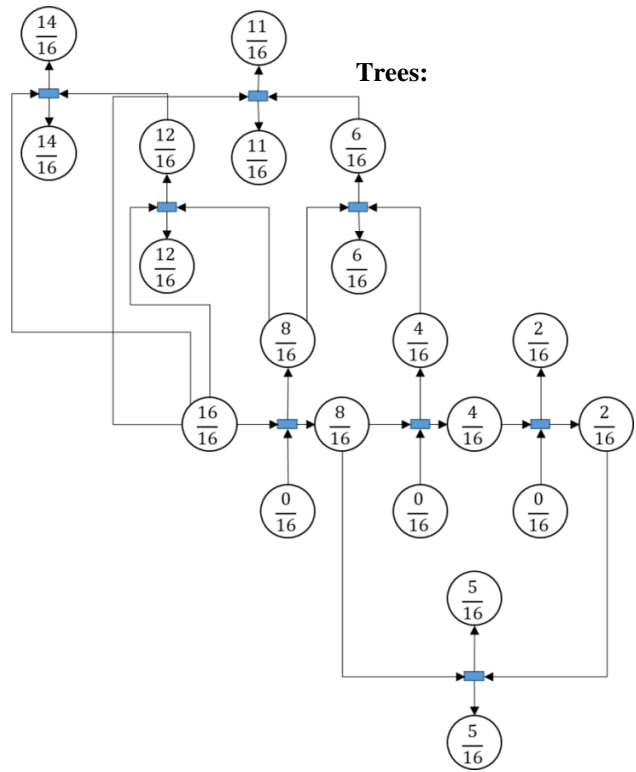

$$TS_1 = \{\frac{5}{16}, \frac{11}{16}, \frac{14}{16}, \frac{16}{16}, \frac{14}{16}, \frac{11}{16}, \frac{5}{16}\}$$

Sample: 5, Buffer: 4, Waste: 2, Steps: 8

**Explanation of Dilution Tree 1:**

An example is shown in the above figure, where the required CF values are as follows:

$\{\frac{5}{16}, \frac{11}{16}, \frac{14}{16}, \frac{16}{16}, \frac{14}{16}, \frac{11}{16}, \frac{5}{16}\}$

Initially, the number of sample droplet=0, the number of buffer droplet=1.

1. In the required CFs, the first required droplet concentration 5/16, is not equal to the sample concentration i.e. 16/16 (100%). Initially storage is empty.

   So, {16/16, 0/16} mixed split operation is executed until an intermediate higher CF droplet is found. 8/16, 4/16, 4/16 are stored in the storage for further use. The series can be generated with two sample droplet and two buffer droplets in two steps. Now the immediate higher droplet of 5/16 is 8/16

2. According to the algorithm 2*5/16 – 8/16 = 2/16 is another CF, which is required to create 5/16 and it is not present in the storage.

   So, it needs to create with two sample procedure. The immediate higher CF of 2/16 is 4/16 which is available in the storage. Which is equal to the twice of 2/16. So, a buffer droplet will be used to dilute the immediate buffer droplet and produce 2/16.

3. {2/16, 8/16} are mix split to produce 5/16 using one step. Now the extra unused droplets are stored within the storage. Hence 2/16 is in the storage at present.

4. Now the next CF value, 11/16, need to be generated. {8/16, 4/16} are mix split to produce 6/16 in one step. Now the extra one unused droplet 6/16 is stored in the storage.

   Now, another droplet 6/16 is there. We mix split {16/16, 6/16} to produce 11/16 in one step.

5. In this way the remaining droplets will be generated.

   Finally, the total dilution process is complete in eight steps with 5 sample droplets and 4 buffer droplets.

**Dilution** 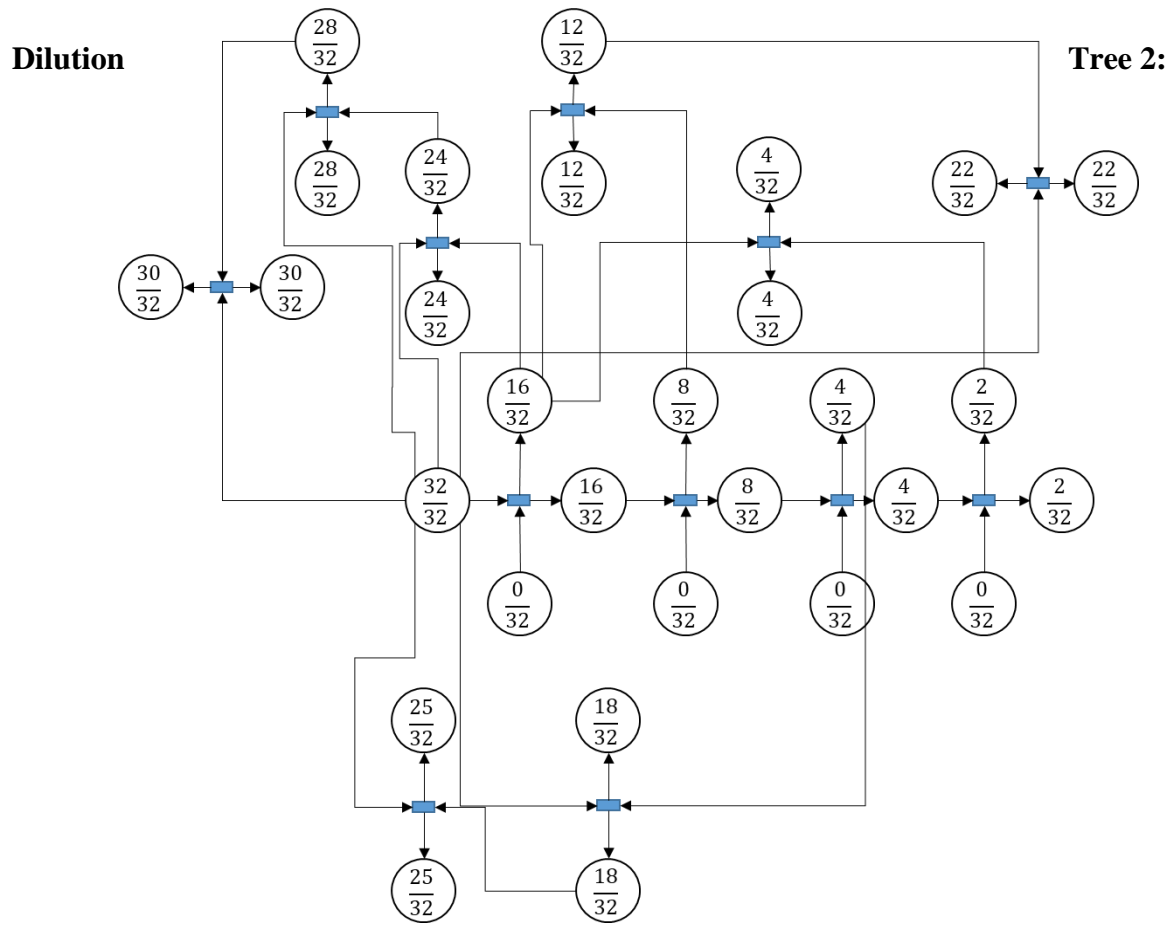 **Tree 2:**

$$TS_2 = \{\frac{9}{32}, \frac{18}{32}, \frac{25}{32}, \frac{30}{32}, \frac{32}{32}, \frac{28}{32}, \frac{22}{32}\}$$

Sample: 8, Buffer: 5, Waste: 4, Steps: 12

Dilution Tree 3:

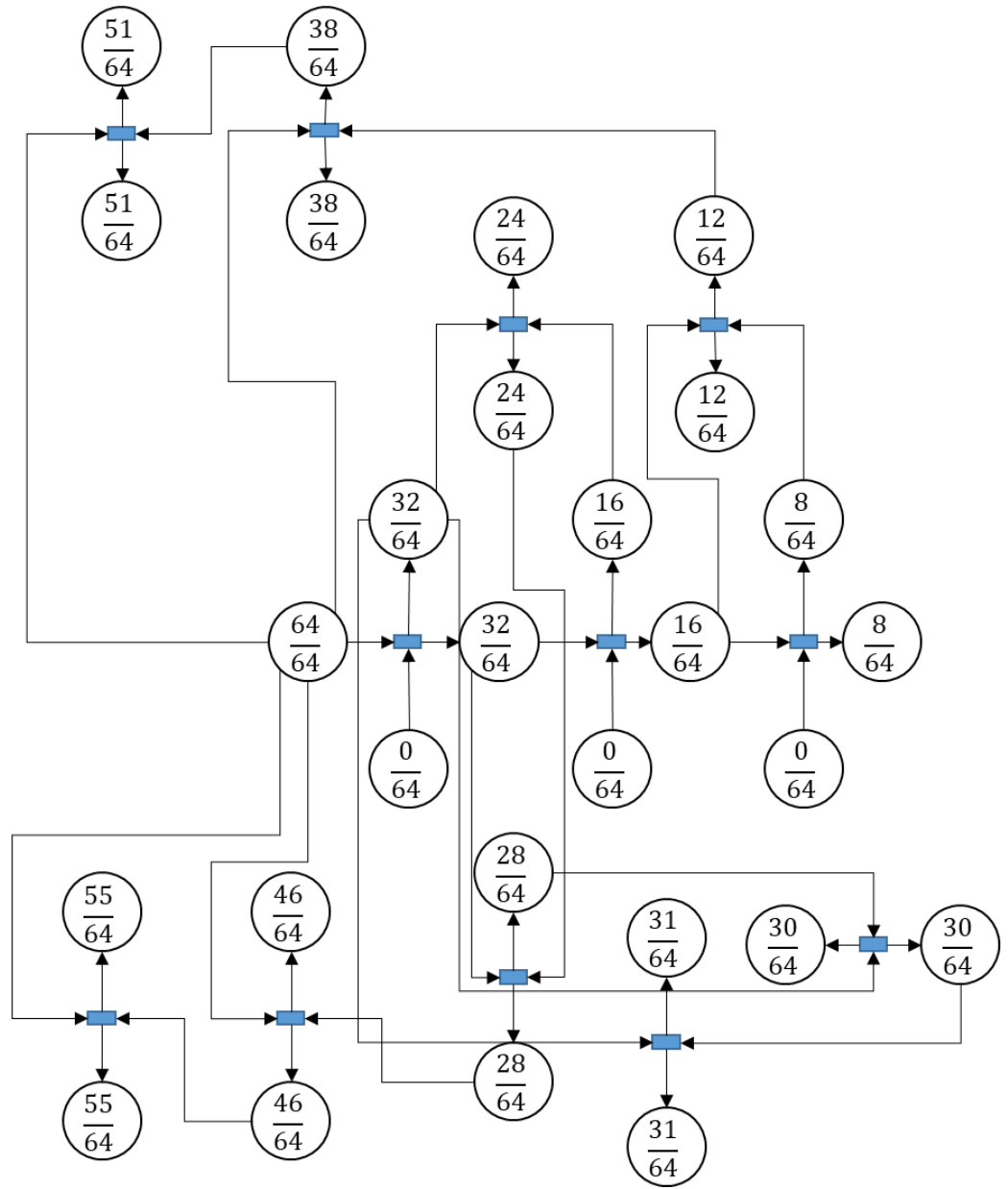

TS$_3$= {$\frac{31}{64}, \frac{55}{64}, \frac{64}{64}, \frac{51}{64}$}

Sample: 6, Buffer: 4, Waste: 8, Steps: 12

# 5. Result and discussion

Here in table 1 we presents the computed result of EMDP algorithm. In this section, we have described the results of multi gradient dilution preparation with EMDP algorithm. There we have reported, the series of dilutions $TS_1 \rightarrow Ts_{10}$ and use have shown the results is table 1. Here we have found that, for the test series $TS_1$, it requires the numbers of sample droplets are 5, number of buffer droplets are 4 and the number of waste droplets are 2. Similarly for $TS_2 \rightarrow TS_{10}$. We have shown the results in table 1

**Table 1: The test series for multi gradient dilution preparation with EDMP algorithm**

| Test Series | # Waste (W) | | | | | | |
| --- | --- | --- | --- | --- | --- | --- | --- |
| | RTWM | REMIA | WARA | RSM | MTC | SWDM | EMDP |
| TS1 | 14 | 6 | 6 | 8 | 8 | 3 | 2 |
| TS2 | 9 | 2 | 2 | 11 | 12 | 5 | 4 |
| TS3 | 14 | 4 | 4 | 10 | 10 | 6 | 8 |
| Test Series | #Sample (S) | | | | | | |
| | RTWM | REMIA | WARA | RSM | MTC | SWDM | EMDP |
| TS1 | 15 | 8 | 8 | 7 | 8 | 6 | 5 |
| TS2 | 17 | 11 | 11 | 9 | 11 | 7 | 8 |
| TS3 | 17 | 6 | 6 | 7 | 9 | 6 | 6 |
| Test Series | #Buffer(B) | | | | | | |
| | RTWM | REMIA | WARA | RSM | MTC | SWDM | EMDP |
| TS1 | 9 | 5 | 3 | 3 | 6 | 4 | 4 |
| TS2 | 10 | 6 | 5 | 5 | 8 | 5 | 5 |
| TS3 | 8 | 3 | 3 | 4 | 3 | 4 | 4 |

**Table 2: Comparative study of different algorithm RTWM, REMIA, WARA, RSM and MTC.**

| Test Series | $CF_s$ | S | B | W | Steps |
| --- | --- | --- | --- | --- | --- |
| $TS_1$ | $\{5/16, 11/16, 14/16, 16/16, 14/16, 11/16, 5/16\}$ | 5 | 4 | 2 | 8 |
| TS$_2$ | $\{9/32, 18/32, 25/32, 30/32, 32/32, 28/32, 22/32\}$ | 8 | 5 | 4 | 12 |
| TS$_3$ | $\{31/64, 55/64, 64/64, 51/64\}$ | 6 | 4 | 8 | 12 |

## 5. Conclusion

In our proposed algorithm EMDP, it can be used for sample and waste aware biochemical on chip protocol .In these method , the prime requirement is the recycling of waste droplets from storage for successful multigradient dilution preparation . The major contribution of our proposed algorithm is to minimize the sample droplets as well as the waste droplets to generate different multigradient sample preparation .We have shown the result for different test series Ts$_1$-Ts$_{10}$. In comparison to the others, the costly sample requirement is much less and also the lesser amount of waste it produces and that is why the cost of the sample droplets is minimized. Here we have compared the result between multi –gradient dilution preparation algorithm and the proposed EMDP. The comparative study shows that overall nearby 50% minimization in the use of sample droplets and waste droplets are achieved using the proposed EMDP. Hence it can be concluded that the proposed EMDP algorithm. May be used for lose cost sample preparation for on-chip bio-chemical process.